\documentclass[epj]{svjour}

\usepackage{amsmath,amsfonts}

\def\hh{\hspace*{10mm}}
\def\hhh{\hspace*{15mm}}

\def\nn{\nonumber}

\def\bx{\boldsymbol{x}}
\def\rd{{\rm d}}

\begin{document}

\title{On the NCCS model of the quantum Hall fluid}

\author{M. Eliashvili\thanks{e-mail: simi@rmi.acnet.ge} \and G. Tsitsishvili
}

\institute{Department of Theoretical Physics,
A. Razmadze Mathematical Institute,
M. Aleksidze 1, Tbilisi 0193 Georgia}

\date{Received: date / Revised version: date}

\abstract{
Area non-preserving transformations in the non-commutative plane are introduced
with the aim to map the $\nu=1$ integer quantum Hall effect (IQHE) state on the
fractional quantum Hall effect (FQHE) $\nu=\frac{1}{2p+1}$ FQHE states.
Using the hydrodynamical description of the quantum Hall fluid, it is shown that
these transformations are generated by vector fields satisfying the Gauss law
in the interacting non-commutative Chern--Simons gauge theory, and the corresponding
field-theory Lagrangian is reconstructed. It is demonstrated that the geometric
transformations induce quantum-mechanical non-unitary similarity transformations,
establishing the interplay between integral and fractional QHEs.
}

\maketitle

\noindent
The apparent similarity between integral and fractional quantum Hall effects challenges
the search for the theoretical schemes that bridge the gap between them.

One of the physically appealing models is the Jain composite fermion (CF) picture
\cite{jain89}, which links the FQHE of interacting electrons to the IQHE of composite
fermions. The CF picture gains mathematical content in the Chern--Simons (CS) gauge theory:
the principal role of the CS gauge field is to attach an even number of elementary magnetic
flux quanta to each electron, converting it to a composite particle.

In the present note we argue that the CS vector potential generates a geometric map between
areas occupied by quantum fluids of different densities. The quantum-mechanical outcomes of
these geometric transformations are the operator similarity transformations
relating the quantum characteristics of the integral and fractional QHE states.

Below we consider QHE states, constituting  the so called  Laughlin series, characterized
by the filling factors:
\begin{equation}
\nu=\frac{n_e}{n_B}=\frac{1}{2p+1},\hh p=0,1,2,... \label{eq:ff}
\end{equation}
Here
\begin{equation}
n_e=\lim_{N,\Omega\rightarrow\infty}\frac{N}{\Omega}
\end{equation}
is the average electron density ($N$ is the number of electrons, and $\Omega$ is the occupied
2-dimensional area), and $n_B=\frac {B}{2\pi}$ is the density of quantum states per Landau
level.

The most important common feature of the $p=0$ and $p\geq 1$ states is that all of them
are incompressible quantum fluids comprising the lowest Landau level (LLL) electrons.
In the case of IQHE the ground state is given by the Slater determinant ($a=1,2,...,N$)
\begin{equation}
\Psi_0\sim
\prod_{a<b}(z_a-z_b){\rm e}^{-\frac{B}{4}\sum_a|z_a|^2} \label{eq:0wf}
\end{equation}
(complex coordinates and the symmetric gauge are used).

At the fractional values of $\nu<1$, which correspond to a partially filled LLL,
the ground state is given by the Laughlin  wave function \cite{laughlin83}
\begin{equation}
\Psi_p\sim
\prod_{a<b}(z_a-z_b)^{2p}\Psi_0. \label{eq:lwf}
\end{equation}
Note  an important detail: the wave functions (\ref{eq:0wf}) and
(\ref{eq:lwf}) both correspond to $N$-particle systems exposed to
one and the same magnetic field $B$, but occupying different areas:
\begin{equation}
\Omega=\frac{2\pi}{B}N\nn
\end{equation}
and
\begin{equation}
\Omega'=(2p+1)\Omega,\label{eq:area1}
\end{equation}
respectively.

The many-particle wave functions (\ref{eq:0wf}) and (\ref{eq:lwf}) satisfy the LLL condition
\begin{equation}
{\hat{\pi}}_{\bar z}(a)\Psi_{\rm LLL}\equiv-{\rm i}\Big[\bar\partial_a+\frac{B}{4}z_a\Big]
\Psi_{\rm LLL}(\bx_1,...,\bx_N)=0. \label{eq:lll}
\end{equation}
Equation (\ref{eq:lll}) may be interpreted as the quantum counterpart of the classical
second class Dirac constraints \cite{dirac64}:
\begin{equation}
\pi_i(a)\equiv p_i(a)+A_i(\bx_a)\approx 0, \label{eq:constr1}
\end{equation}
where $p_i$ are the canonical momenta, and $A_i(\bx)=\frac{B}{2}\epsilon_{ik}x_k$ is the
electromagnetic vector potential.

The constraints (\ref{eq:constr1}) may be found from the zero-mass Lagrangian \cite{dunne90}
\begin{equation}
L=-\sum_{a=1}^N\Big[\dot x_i(a,t)\frac{B}{2}\epsilon_{ik}x_k(a,t)\Big].\label{eq:0lag}
\end{equation}
We do not include the confining potential, assuming that it takes some constant value
in $\Omega$ and affects only edge states. The corresponding classical dynamical equations
are given by
\begin{equation}
\dot x_i(a,t)=0.  \label{eq:de}
\end{equation}
In other words, classically the electrons are frozen, i.e. they occupy fixed positions:
\begin{equation}
x_{0i}(a,t)=a_i. \label{eq:fix}
\end{equation}

Assuming that the system behaves like a perfect fluid one can pass to the hydrodynamical
description \cite{susskind01}, i.e. one may consider the electron system as a continuous
distribution of particles, occupying the area
\begin{equation}
\Omega=\int_D\rd^2x.\label{eq:area}
\end{equation}
Particles are labeled by a continuous variable -- the co-moving
coordinate $\boldsymbol{\xi}$, which is introduced via the replacement
$\bx(a,t)\rightarrow \bx(\boldsymbol{\xi},t)$. The Lagrange variables
{$\boldsymbol{\xi}$} are fixed by the conditions
\begin{equation}
\xi_i=x_i(\boldsymbol{\xi},0). \label{eq:cmc}
\end{equation}

The zero-mass hydrodynamical Lagrangian is given by
\begin{equation}
\mathcal L=\int_D\rd^2\xi\rho_0\Big[-\frac{B}{2}\epsilon_{ik}
\dot x_i(\boldsymbol{\xi},t)x_k(\boldsymbol{\xi},t)\Big], \label{eq:hl1}
\end{equation}
and we suppose that (\ref{eq:hl1}) corresponds to the IQHE state with filling factor
$\nu=1$ and with a constant density, $\rho_0=n_B$.

In accord with (\ref{eq:fix}), for the real trajectories we have
\begin{equation}
\bx_0(\boldsymbol{\xi},t)=\boldsymbol{\xi}, \label{eq:euler}
\end{equation}
and one may set $\boldsymbol{\xi}\in D$. Consequently the occupied area is related
to the particle density by the equation
\begin{equation}
\Omega=\int_D\rd^2\xi=\rho_0^{-1}\int_D\rd^2\xi\langle\rho(\boldsymbol{\xi})\rangle,
\label{eq:area2a}
\end{equation}
where $\langle\rho(\boldsymbol{\xi})\rangle$ is the microscopic density corresponding
to $\nu=1$.

Now consider the second droplet occupying the area $\Omega'$, assuming that both systems
contain the same quantity of fluid. The primed system is described by the Lagrangian
\begin{equation}
\mathcal L'=\int_{D'}\rd^2\xi'\rho'_0(\boldsymbol{\xi}')\Big[-\frac{B}{2}\epsilon_{ik}\dot
x'_i(\boldsymbol{\xi}',t)x'_k(\boldsymbol{\xi}',t)\Big] \label{eq:hl2}
\end{equation}
under the conditions that
\begin{align}
\Omega'&=\int_{D'}\rd^2\xi'=(2p+1)\int_{D}\rd^2\xi, \nn\\
N&=\int_{D}\rd^2\xi\rho_0=N'=\int_{D'}\rd^2\xi'\rho'_0(\boldsymbol{\xi}').\label{eq:cond1}
\end{align}
In other words, the Lagrangian (\ref{eq:hl2})  corresponds to
$\nu=\frac{1}{2p+1}$.

The transition to the primed system is realized by means of the
map $D\rightarrow D'$ \cite{eliashvili00,eliashvili99}:
\begin{equation}
\xi_i\rightarrow\xi'_i=F_i(\xi)=\xi_i+\theta\epsilon_{ik}f_k(\xi,0),\hspace*{5mm}
\theta=-\frac{1}{B}. \label{eq:map1}
\end{equation}
In (\ref{eq:map1}) $f_k(\xi,t)$ is a time-dependent vector field
(here and below vector indices are omitted when they are obvious).

The hydrodynamical variables of the primed system are defined as follows:
\begin{align}
x_i'(\xi',t)&=x_{0i}(\xi,t)+\theta\epsilon_{ik}f_k(\xi,t),\hh x_{0i}(\xi,t)\equiv\xi_i\nn\\
\dot x_i'(\xi',t)&=\theta\epsilon_{ik} \dot f_k(\xi,t),\label{eq:map2}
\end{align}
and the corresponding Lagrangian and the occupied area are given by
\begin{equation}
\mathcal L'=\int_{D}\rd^2\xi\rho_0\Big[\frac{1}{2B}\epsilon_{ik} f_i(\xi,t)\dot
f_k(\xi,t)\Big] \label{eq:hl3}
\end{equation}
and
\begin{equation}
\Omega'=\int_D\rd^2\xi\big[1+\epsilon_{ik} F_{ik}(\xi,t)\big],\label{eq:area'}
\end{equation}
respectively.

In the latter expression
\begin{align}
F_{ik}(\xi,t)=D_if_k(\xi,t)&\equiv\partial_i f_k+\frac12\{f_i,f_k\}_{\rm D}\nn\\
&\equiv\partial_if_k+\frac12\theta\epsilon_{mn}\frac{\partial f_i}{\partial\xi_m}
\frac{\partial f_k}{\partial\xi_n},\label{eq:rotc}
\end{align}
and taking into account the definition of the filling factor one gets
\begin{equation}
\int_D\rd^2\xi \epsilon_{ik} F_{ik}(\xi,t)=2p\Omega. \label{eq:r2}
\end{equation}
Now recall (\ref{eq:area2a}) and assume that one may convert the integral
constraint (\ref{eq:r2}) to the local equation
\begin{equation}
4\pi p\langle\rho(\xi,t)\rangle+\epsilon_{ik}D_if_k(\xi,t)=0.\label{eq:GL}
\end{equation}
Introducing the Lagrange multiplier $f_0(\xi,t)$, the  constraint
(\ref{eq:GL}) may be combined with (\ref{eq:hl3}), resulting in the
Lagrangian
\begin{align}
\mathcal L_{\rm S}&=\int_{D}\rd^2\xi\rho_0\bigg\{\frac{1}{2B}\epsilon_{ik}
f_i(\xi,t)\dot f_k(\xi,t)+\nn\\
&+\theta f_0(\xi,t)\Big[4\pi
p\langle\rho(\xi,t)\rangle+\epsilon_{ik} D_if_k(\xi,t)\Big]\bigg\}.  \label{eq:hl4}
\end{align}

In the thermodynamic limit ($N\rightarrow\infty$, $\Omega\rightarrow\infty$)
the Lagrangian (\ref{eq:hl4}) is equivalent to
\begin{align}
\mathcal L_{\rm SCS}=\int_{R^2}\rd^2\xi\Big[\langle\rho(\xi,t)\rangle f_0(\xi,t)+\nn\\
+\frac{\kappa}{2}\epsilon^{\mu\nu\lambda}f_\mu(\xi,t)
D_\nu f_\lambda(\xi,t)\Big],  \label{eq:hl4a}
\end{align}
where the covariant curl is
\begin{equation}
D_\mu f_\nu=\partial_\mu f_\nu+\frac12\{f_\mu,f_\nu\}_{\rm D}, \hspace*{5mm}
(\mu,\nu=0,1,2).
\end{equation}

Following Susskind \cite{susskind01}, the symplectic CS Lagrangian
$\mathcal L_{SCS}$ (with the zero source term) may be interpreted
as a truncation of the non-commutative Chern--Simons (NCCS) Lagrangian.

In what follows we will directly get the NCCS Lagrangian considering geometric mappings
in the non-commutative space. Non-commutativity enters on replacing the Poisson brackets
of two canonical variables by the Dirac brackets \cite{dirac64} generated by the constraints
(\ref{eq:constr1}). In particular, the Dirac bracket
\begin{equation}
\{x(a),y(b)\}_{\rm D}=\theta\delta_{ab}=-\frac{1}{B}\delta_{ab}\label{eq:db}
\end{equation}
leads to non-commuting coordinate operators:
\begin{equation}
[\hat x_i(a),\hat x_k(b)]=i\epsilon_{ik}\theta\delta_{ab}. \label{eq:ncc}
\end{equation}
Notice that, due to the non-commutativity of the coordinates, the hydrodynamical variables $\xi_i$
have to be replaced by non-commutative quantities and the Lagrangian (\ref{eq:hl1}) by its NC
analogue,
\begin{equation}
\mathcal L_{\rm NC}=\int_D\rd^2\xi\rho_0\Big[-\frac{B}{2}\epsilon_{ik}\dot x_i(\xi,t)\star
x_k(\xi,t)\Big],\label{eq:l2}
\end{equation}
where
\begin{equation}
f(\xi)\star g(\xi )={\rm e}^{\frac{{\rm i}}{2}\theta\epsilon_{ik}\partial_i\partial'_k}
f(\xi)\cdot g(\xi')\vert_{\xi'=\xi} \label{eq:gm}
\end{equation}
is the {Groenewold-Moyal} star product.

Correspondingly, instead of the transformation (\ref{eq:map1}) one has to consider the
operator homomorphism \cite{eliashvili03}
\begin{equation}
\hat{\mathcal W}[\xi_i]=\hat\xi_i\rightarrow \hat{\mathcal W}[F_i]. \label{eq:qmap}
\end{equation}
Here, by
\begin{equation}
\hat{\mathcal W}[F_i]=\frac{1}{(2\pi)^2}\int \rd^2p\int
\rd^2x{\rm e}^{-{\rm i}p_i(\hat \xi_i-x_i)}F_i(x)\label{eq:Weyl}
\end{equation}
is denoted the symbol of Weyl ordering.

The resulting Lagrangian for the primed system will be
\begin{equation}
\mathcal L'_{\rm NC}=\int_{D}\rd^2\xi\rho_0\Big[\frac{1}{2B}\epsilon_{ik}
f_i(\xi,t)\star\dot f_k(\xi,t)\Big], \label{eq:hl5}
\end{equation}
where the field $f_i(\xi,t)$ has to satisfy the NC analogue of (\ref{eq:r2}), with
\begin{align}
F_{ik}(\xi,t)&=\theta\int_{R^2}\rd^2xD(\xi-x)[\partial_if_k(x)-{\rm i}f_i(x)\star
f_k(x)]\nn\\
&\equiv\theta\int_{R^2}\rd^2xD(\xi-x)f_{ik}(x,t). \label{eq:rotnc}
\end{align}
Here
\begin{equation}
D(\xi-x)=\frac{1}{\pi\theta}{\rm e}^{-\frac{1}{\theta}(x_i-\xi_i)^2},
\end{equation}
and in the commutative limit (\ref{eq:rotnc}) reduces to (\ref{eq:rotc}).
The last expressions may be derived by taking into consideration the area transformation
rule in the NC plane \cite{eliashvili03}.

The NC analogue of the constraint equation (\ref{eq:r2}) is given by
\begin{align}
2p\Omega&=\frac{4\pi p}{B}\int_D\rd^2\xi\langle\rho(\xi)\rangle\nn\\
&=-\frac{1}{B}\int_D \rd^2\xi\int_{R^2}\rd^2xD(\xi-x)\epsilon_{ik} f_{ik}(x,t)\label{eq:GLconst}
\end{align}

Recall, that the density
\begin{equation}
\langle\rho(\xi,t)\rangle:=\int_{R^2}\rd^2x D(\xi-x)\rho(x,t) \label{eq:dens}
\end{equation}
corresponds to $\nu=1$. The non-commutative version of the Gauss law (GL) looks like
\begin{equation}
4\pi p\rho(x,t)+\epsilon_{ik}f_{ik}(x,t)=0. \label{eq:GL1}
\end{equation}
Introducing the Lagrange multiplier $f_0(\xi,t)$ for the constraint (\ref{eq:GL1}), we
arrive at the Lagrangian
\begin{equation}
\mathcal L_{\rm NC}=4\pi
p\theta\rho_0\int_D\rd^2\xi\Big[\rho(\xi) f_0(\xi)-\frac{\kappa}{2}
\epsilon_{ik}\big( f_i\star \dot f_k-f_0 f_{ik}\big)\Big], \label{eq:hl6}
\end{equation}
where  $k^{-1}=4\pi p$. In the thermodynamical limit (\ref{eq:hl6}) is
equivalent to the NCCS Lagrangian
\begin{align}
\mathcal L_{\rm NCCS}&=\int_{R^2}\rd^2 \xi\bigg[\rho(\xi,t)\star f_0(\xi,t)\nn\\
&+\frac{\kappa}{2}\varepsilon^{\mu\nu\lambda}
f_\mu\star\bigg(\partial_\nu f_\lambda-{\rm i}\frac23f_\nu\star f_\lambda\bigg)\bigg].\label{eq:NCCS}
\end{align}

Some comments are in order here. As we  have already remarked, in the context of QHEs
this kind of Lagrangian was first introduced in \cite{susskind01} on the basis of area
preserving diffeomorphisms (APDs) in the commutative plane. In our case the underlying
transformations realize mappings between different areas in the non-commutative plane,
i.e. they belong to the class of area non-preserving transformations. Non-commutative
APDs are represented by the NC
gauge transformations
\begin{equation}
f_i(\xi,t)\rightarrow f_i(\xi,t)+\frac{{\rm i}}{2\theta}\epsilon_{ik}(\xi_k\star
\lambda-\lambda\star\xi_k), \label{eq:gaugetr}
\end{equation}
under which the GL (\ref{eq:GL1}) is invariant.

Hence, the area transformation rule in the non-commutative plane leads to the GL in the
NCCS gauge theory. In complex notation ($z=\xi_1+{\rm i}\xi_2$) it looks like
\begin{equation}
\Big(\bar\partial f_z-\partial
f_{\bar z}\Big)+ \big(f_z \star f_{\bar z}-f_{\bar z}\star f_z\big )=2{\rm i}\pi
p \rho(\xi).
\end{equation}
This non-linear equation simplifies in the holomorphic gauge $f_{\bar z}=0$, reducing
the GL to the equation
\begin{equation}
\bar\partial f_z(\xi)=2{\rm i}\pi p\rho(\xi). \label{eq:holg}
\end{equation}
The solution to (\ref{eq:holg}) is given by
\begin{equation}
f_z(\xi)=2{\rm i}p\partial\int\rd^2\xi'\ln(z-z')\rho(\xi')={\rm i} S^{-1}\partial S,\label{eq:fz}
\end{equation}
where the holomorphic function
\begin{equation}
S={\rm e}^{2p\int\rd^2\xi'\ln(z-z')\rho(\xi')},\hh \bar\partial S=0. \label{eq:S}
\end{equation}

One easily verifies that the geometric transformation (\ref{eq:map1}) induces the operator
homomorphism
\begin{align}
W(\xi_i)\rightarrow W(\xi_i')&=W(S^{-1})W(\xi_i)W(S)=\nn\\
\nn\\
&=W(\xi_i+\theta\epsilon_{ik}f_k(\xi)). \label{eq:optr1}
\end{align}
The expression (\ref{eq:optr1}) is the non-commutative space analogue of the usual coordinate
transformation. The latter may be associated with a transformation of the mean values:
\begin{align}
\langle\Phi|W(\xi_i)|\Phi\rangle&\rightarrow\langle\Phi|W(\xi'_i)|\Phi\rangle=\nn\\
\nn\\
&=\langle\Phi|W(S^{-1})W( \xi_i)W(S)|\Phi\rangle.\label{eq:tr1}
\end{align}

As an alternative, one may attribute geometric transformations to the map in the Hilbert space
$|\Phi\rangle\rightarrow|\Phi'\rangle=W(S)|\Phi\rangle$,
$\langle\Phi|\rightarrow\langle\Phi|W(S^{-1})$, keeping the coordinate operators  unchanged:
$W(\xi_i)\rightarrow W(\xi_i)$.

In the NC plane the operators $\hat z$ and $\hat{\bar z}$ are realized by
\begin{equation}
W(z)\equiv \hat z=z,\hhh W(\bar z)\equiv\hat{\bar z}
=\frac{2}{B}\frac{\partial}{\partial z}\nn
\end{equation}
and
\begin{equation}
W(S)=S.
\end{equation}
Then the operator transformations
\begin{align}
S^{-1}\hat{\bar z}S&=\hat{\bar z}+2{\rm i}\theta f_z(z)\nn\\
S^{-1}\hat{z}S&=\hat z \label{eq:optr}
\end{align}
reproduce in operator form the map $\xi_i\rightarrow\xi_i'$.

Now one may go back to the quantum-mechanical picture. The reference, i.e. the $\nu=1$
QHE state, is described by the wave function (\ref{eq:0wf}) and the dynamics is governed
by the constraint equations (\ref{eq:lll}). Assuming that for the microscopic density
one has
\begin{equation}
\rho(\bx)=\sum_a\delta(\bx-\bx_a),
\end{equation}
one gets the result that the geometric transformation is realized by the vector field
\begin{equation}
f_z(\bx_a)=2{\rm i}p{\sum_{b}}'\frac{1}{z_a-z_b},\hh f_{\bar z}(\bx_a)=0.
\end{equation}
The holomorphic function
\begin{equation}
S_p(z_1,...,z_n)={\rm e}^{2p\sum_{a<b}\ln(z_a-z_b)}=\prod_{a<b}(z_a-z_b)^{2p}
\end{equation}
generates the transformations
\begin{align}
z_a&\rightarrow z_a,\nn\\
\\
\frac{\partial}{\partial z_a}&\rightarrow S_p^{-1}
\frac{\partial}{\partial z_a}S_p=\frac{\partial}{\partial z_a}
+2p{\sum_b}'\frac{1}{z_a-z_b}.\nn
\end{align}

In the alternative picture the transition from the $\nu=1$ IQHE state  to the
$\nu=\frac{1}{2p+1}$ FQHE state  is accomplished by the map
\begin{equation}
\Psi_p=
\Psi_0\rightarrow S_p\Psi_0=\prod_{a<b}(z_a-z_b)^{2p+1}
{\rm e}^{-\frac{B}{4}\sum_a|z_a|^2},\label{eq:Sim1}
\end{equation}
reproducing the Laughlin wave function (\ref{eq:lwf}). In parallel, the relevant quantum
operators (like constraints or guiding center coordinates) have to undergo the similarity
transformations
\begin{equation}
\hat{\mathcal O}_0\rightarrow\hat{\mathcal O}_p
=S_p \hat{\mathcal O}_0 S_p^{-1}.\label{eq:Sim2}
\end{equation}

In particular, the LLL constraint remains invariant:
\begin{equation}
\hat\pi_{\bar z}(a)\rightarrow \hat\Pi_{\bar z}(a)
=S_p\hat\pi_{\bar z}(a)S_p^{-1}=\hat\pi_{\bar z}(a),
\end{equation}
and the LLL condition is preserved,
\begin{equation}
\hat\pi_{\bar z}\Psi_p=0.\label{lll1}\\
\end{equation}

Note that the similarity  transformations (\ref{eq:Sim1}) and (\ref{eq:Sim2}) are
induced by geometric mappings relating different quantum Hall droplets and they
establish a non-unitary equivalence between the integral $(p=0)$ and fractional
$(p>1)$ quantum Hall states.

\vspace*{5mm}

\noindent {\it Acknowledgements.}
M.E. thanks P. Sorba for stimulating discussions. G.T. is grateful to the Associate
Programme of the Abdus Salam ICTP where part of this work was performed.
The work was in part supported by the Georgian National Science
Foundation under  grants GNSF/ST-06/4-018 and GNSF/ST-06/4-050

\end{document}